**Super-Droplet-Repellent Carbon-Based Printable Perovskite Solar Cells**

*Cuc Thi Kim Mai\*, Janne Halme\*, Heikki A. Nurmi, Aldeliane M. da Silva, Gabriela S. Lorite, David Martineau, Stéphanie Narbey, Naeimeh Mozaffari, Robin H. A. Ras, Syed Ghufran Hashmi and, Maja Vuckovac\**

Cuc Thi Kim Mai, Aldeliane M. da Silva, Gabriela S. Lorite, Syed Ghufran Hashmi
Microelectronics Research Unit, Faculty of Information Technology & Electrical Engineering, P. O. Box 8000, FI-90014, University of Oulu, Finland.
E-mail: Thi.Mai@oulu.fi

Janne Halme, Heikki A. Nurmi, Robin H. A. Ras, Maja Vuckovac,
Department of Applied Physics, Aalto University School of Science, PO Box 11000, FI-00076, Finland.
E-mail: janne.halme@aalto.fi, maja.vuckovac@aalto.fi

Heikki A. Nurmi, Robin H. A. Ras, Maja Vuckovac
Centre of Excellence in Life-Inspired Hybrid Materials (LIBER), Aalto University, Espoo, Finland.
E-mail: maja.vuckovac@aalto.fi

David Martineau, Stéphanie Narbey
Solaronix SA, Rue de l' Ouriette 129, CH-1170 Aubonne, Switzerland

Naeimeh Mozaffari
Department of Materials Science and Engineering, Monash University, Clayton, Victoria 3800, Australia

**Abstract**

Despite attractive cost-effectiveness, scalability, and superior stability, carbon-based printable perovskite solar cells (CPSCs) still face moisture-induced degradation that limits their lifespan and commercial potential. Here, we investigate the moisture-preventing mechanisms of thin nanostructured super-repellent coating (advancing contact angle > 167° and contact angle hysteresis 7°) integrated into CPSCs for different moisture forms (falling water droplets vs water vapor vs condensed water droplets). We show that unencapsulated super-repellent CPSCs have superior performance under continuous droplet impact for 12h (rain simulation experiments) compared to unencapsulated pristine (uncoated) CPSCs that degrade within seconds. Contrary to falling water droplets, where super-repellent coating serves as a shield, we found water vapor to physisorb through porous super-repellent coating (room temperature and relative humidity, RH 65% and 85%) that increased the CPSCs performance for 21% during ~43 days similarly to pristine CPSCs. We further showed that, water condensation forms within or below the super-repellent coating (40˚C and RH 85%), followed by chemisorption and degradation of CPSCs. Because different forms of water have distinct effect on CPSC, we suggest that future standard tests for repellent CPSCs should include rain simulation and condensation tests. Our findings will thus inspire the development of super-repellent coatings for moisture prevention.

## 1. Introduction

Solution-processed perovskite solar cells (PSCs) are advancing rapidly, with a remarkable increase in power conversion efficiency (PCE) from 3.8 %[1] to 26.1 %.[2] The low-cost and scalable production methods make PSCs one of the most promising photovoltaic (PV) technologies, especially in electricity generation at grid scales[3–7] and possible integration into light-harvesting applications for indoor building environments.[8–10] However, current PSCs lack long-term operational stability under various environmental conditions, preventing them from reaching their full commercial potential.[11–13] This is primarily due to hybrid metal halide perovskites light absorbers (among other active layers) that rapidly degrade when exposed to environmental factors, particularly moisture, impairing their PV performance and stability.[14–19]

An effective way to prevent moisture is to isolate the device from the surrounding environment by encapsulation[20–23] or implementing moisture barrier layers.[24,25] The most common encapsulation method is the so-called glass-glass method, where the device is sandwiched between two glasses with an adhesive encapsulant and sealed around with sealant (Kapton,[26] butyl rubber,[27,28] UV glue,[29] epoxy resin,[30–32] paraffin,[33] or polyisobutylene (PIB).[34] So far, this method has been an effective strategy in preventing moisture degradation of PSCs. However, complex packaging makes the process time-consuming, highly costly, and challenging for sealing large areas.[35] Additionally, it is incompatible with flexible devices.[35]

Here, we discuss the concept of hydrophobicity in the passivation process[36] introduced among other moisture barrier layers (Table S1). Due to their low surface energy, hydrophobic coatings reduce the contact area between water droplets and solid surfaces. In other words, low surface energy restricts the spreading of the droplets, resulting in droplets forming hemispherical or spherical shapes with contact angles between 90° and 150° (90° < $\theta$ < 150°).[37] However, droplets on such coatings could experience significant friction and adhesion forces, resulting in sticky or immobile droplets.[38] Such droplets, over time, can damage the hydrophobic coating, allowing moisture to penetrate and further degrade PSCs (serves as a wetting defect, and water has a higher affinity to accumulate at the defect). This can be addressed by reducing contact points between droplet and surface using superhydrophobic (water super-repellent) coatings that combine roughness and low surface energy to achieve advancing contact angle $\theta_{\text{adv}} > 150°$ and contact angle hysteresis CAH < 10 °.[37] The CAH should be essential in PSC applications as it measures droplet mobility and how droplets advance on and recede from the surface. It is usually calculated as the difference between the advancing contact angle ($\theta_{\text{adv}}$, measured by increasing droplet volume) and receding contact angle ($\theta_{\text{rec}}$, measured by decreasing droplet volume). As the smaller CAH is, the droplet is more mobile[38] and will spend less time in contact with the surface, drastically reducing the chance for PSC moisture-induced degradation. Current studies employ rather hydrophobic than superhydrophobic materials[35,39–46] and usually a couple of hydrophobic (e.g., PDMS) barrier layers similar to sealant materials (dense and not only as the top layer)[44] to demonstrate improved stability of PSCs. Surprisingly, there remains a lack of investigation into the effect of the superhydrophobic coatings on PSCs performance. Moreover, the standard aging tests that engage different moisture (water) forms are lacking.

Despite recent developments in PSCs, it remains unclear how the hydrophobic/superhydrophobic coatings improve the performance of PSCs. Discussion on the moisture-preventing mechanisms of those coatings is scarce, and the chosen aging tests easily fall out of ISOS recommendations.[47] Here we show that super-repellent coatings ($\theta_{\text{adv}} > 150°$ and CAH < 10°) shields against bulk water, making them remarkably stable compared to pristine (uncoated) CPSCs. We intentionally chose the colloidal dispersion of hydrophobic silica nanoparticles (commercially available Glaco coating) to highlight the importance of understanding moisture-induced degradation mechanisms (bulk water vs vapor vs condensed droplets). While superior to bulk water protection (rain simulation), we found the super-repellent coating to allow vapor physisorption at room temperature, leading to improvement and condensate formation at elevated temperatures, leading to degradation of CPSCs performances. We thus propose rain simulation and condensation as additional aging tests for hydrophobic/superhydrophobic PSCs. We believe such aging tests can guide coatings, barrier layers, and encapsulation development to prevent moisture degradation of PSCs.

## 2. Results and Discussion

### 2.1. Unencapsulated Super-Repellent Carbon-Based Printable Perovskite Solar Cells

Carbon-based printable perovskite solar cells (CPSCs) used in this work have Glass/FTO/c-TiO$_2$/mp-TiO$_2$/mp-ZrO$_2$/carbon/infiltrated perovskite structure.[48] The compact TiO$_2$ (c-TiO$_2$) layer is deposited using spray pyrolysis, while mesoporous layers of TiO$_2$ (mp-TiO$_2$), ZrO$_2$ (mp-ZrO$_2$), and carbon are fabricated via screen-printing. A perovskite precursor solution of methylammonium lead iodide (MAPbI$_3$) and 5-ammonium valeric acid iodide (5-AVAI) is infiltrated throughout the electrode stack using a programmable multi-channel pipetting robot (Solaronix). The samples are annealed at 55˚C for 90 minutes and then cooled to room temperature to form perovskite crystals in the porous electrode structure (**Figure 1a**). These unencapsulated CPSCs are denoted as pristine CPSCs. Contact angle measurement show their hydrophilic nature ($\theta < 90°$) with advancing, $\theta_{\text{adv}}$ and receding, $\theta_{\text{rec}}$ contact angles being 53˚ and 14˚, respectively (Figure 1b). During contact angle measurements, the pristine CPSC turns yellow, showing immediate degradation when it comes in contact with macroscopic droplets (~8 µL). As expected, scanning electron microscopy (SEM) imaging shows porous carbon nanoparticles and graphite flakes (Figure 1c).

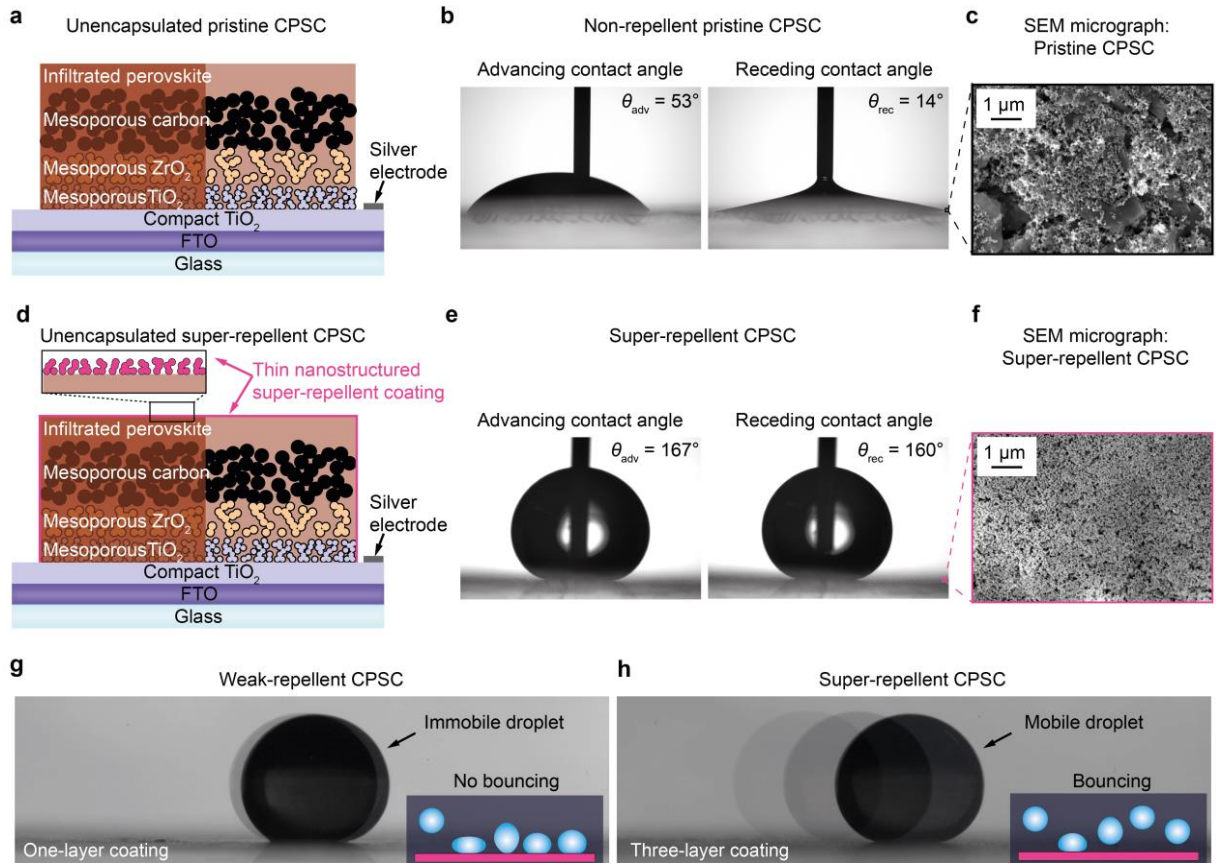

**Figure 1.** Unencapsulated carbon-based printable perovskite solar cells (CPSC): Pristine vs Super-repellent CPSCs. (a) Schematic illustration of pristine CPSC (non-coated) showing the layered stack structure: FTO glass, a thin layer of compact $TiO_2$, and the mesoporous layers ($TiO_2$, $ZrO_2$, and carbon electrode) infiltrated with perovskite with its (b) wetting properties (measured advancing and receding contact angles) and (c) scanning electron microscopy (SEM) micrograph (top view). (d) Schematic illustration of super-repellent CPSC coated with a thin nanostructured super-repellent coating (commercially available Glaco) with its (e) wetting properties (measured advancing and receding contact angles) and (f) SEM micrograph (top view). The inset in (f) shows the porous structure of the coating. (g-h) Video frames captures from oscillation droplet tribometer measurements showing (g) immobile ferrofluid droplet on one-layer coated CPSC indicating no droplet bouncing (weak-repellent CPSC) and (h) highly mobile ferrofluid droplet on three-layer coated CPSC resulted in droplet bouncing.

To achieve super-repellency, we introduce a thin transparent layer (Figure S1) of hydrophobic silica nanoparticles (commercially available Glaco Mirror Coat Zero) (Figure 1d), and these CPSCs are denoted as super-repellent CPSCs. The coating was applied directly on the carbon layer using a spin-coater, and one- vs three-layer coating was explored to achieve the best water repellence. By measuring contact angles, we found a massive improvement in repellence (Figure 1e), as both $\theta_{adv}$ and $\theta_{rec}$ increases drastically to 167° and 160°, while CAH decreased from 39° for pristine CPSC to 7° for super-repellent CPSC. SEM images reveal full and homogeneous coverage with silica nanoparticles (Figure 1f).

We further studied droplet mobility with an oscillating droplet tribometer (ODT)[49,50] that measures friction forces of water-like ferrofluid droplets moving on a repellent surface (Figure 1g). Despite the low CAH (obtained with contact angle measurements) for both one- and three-layer coated CPSCs, the droplets were immobile (pinned) on one-layer coated samples (Video S1), indicating large friction forces due to less or non-uniform coverage of hydrophobic silica nanoparticles, resulting in weakly-repellent CPSCs (Figure 1g). This further implies no droplet bouncing on weakly-repellent CPSCs. On the other hand, the droplet was highly mobile for three-layer coated samples (Video S2), experiencing friction forces of $710 \pm 26$ nN typical for super-repellent surfaces[51,52] (Figure 1h). This indicates sufficient coverage with a three-layer coating that allows droplet bouncing (Video S3). Thus, the following discussion will be based on the unencapsulated three-layer coated super-repellent CPSCs.

## 2.2. Photovoltaic (PV) Performance of Super-Repellent CPSCs

To investigate how super-repellent coating is integrated into CPSC, we examined unencapsulated CPSCs (34 Batch I and 16 Batch II) by measuring their PV parameters (Figure S2) before (pristine) and after applying coating (super-repellent). To provide more reliable and realistic results, we perform current-voltage (J-V) measurements on all CPSCs (an active area of 1.5 cm$^2$) using an aperture area of 0.64 cm$^2$. By reducing the aperture area from 0.64 cm$^2$ to 0.14 cm$^2$, the PCE values of the reverse scan increase from ~8% up to ~11 % (Figure S2). Then we coated the CPSCs with the super-repellent coating (17 from Batch I and 8 from Batch II) and measured their PV parameters after applying the coating. The measured power conversion efficiency (PCE) for pristine CPSCs (before coating) and super-repellent CPSCs (after coating) (17 of each from Batch I) is shown in **Figure 2a**, while all PV parameters are shown in Figure S3.

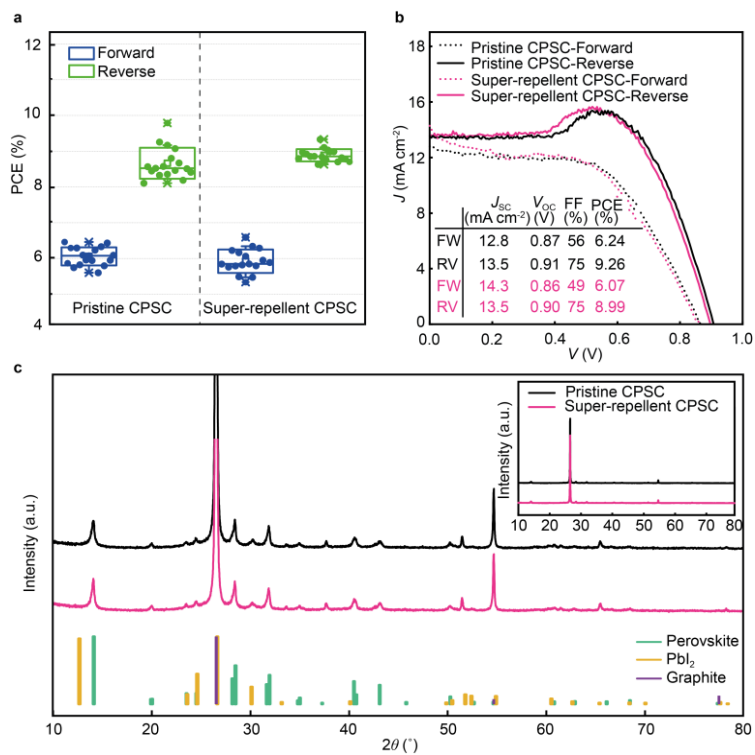

**Figure 2.** Integration of super-repellent coating into CPSC. (a) Statistical PCE for both forward and backward scans for 17 unencapsulated CPSCs before applying coating (pristine) and the same CPSC after applying coating (super-repellent CPSCs) (from Batch I). (b) J-V curves of representative pristine and super-repellent CPSCs with a measured aperture area of 0.64 cm$^2$ under 1 Sun condition. (c) XRD patterns of pristine and super-repellent CPSC. Note that no additional peaks characteristic to super-repellent coating (hydrophobic silica nanoparticles) were found primarily due to the amorphous structure of the coating.

We found average PCE of pristine CPSCs and super-repellent CPSCs to be almost identical (pristine CPSCs is 7.2 % ± 0.3 %, and for super-repellent CPSCs 7.3 % ± 0.2 %), giving no statistically significant differences in the PCE when applying the super-repellent coating (t-test with 95 % confidence) (Table S2). The average photocurrent density ($J_{SC}$) and open circuit voltage ($V_{OC}$) (Figure 2b) showed a statistically significant effect of super-repellent coating with a p-value less than 0.05. However, the effect size was insignificant, with a change of less than 5 % (Table S2). This indicates that the applied super-repellent coating did not cause structural changes to CPSCs. This was confirmed with XRD measurements (Figure 2c) that showed no effect on the perovskite crystal structure as the d-spacing and the intensity of peaks remained constant after coating application. Also, no new PbI$_2$ peak was formed for super-repellent CPSCs. Thus, an excellent integration of super-repellent coating has been achieved. The results from Batch II confirmed that the coating did not impact initial PV performance and showed excellent reproducibility of the application process (Figure S4 and Table S3 in Supporting Information). In this work, we aim not to improve PCE through synthesis but to explore the effect of super-repellent coating on CPSCs performances and how the coating performs in the presence of different moisture forms (falling water droplets vs water vapor vs condensed droplets).

**2.3. Performance of Super-Repellent CPSC in the presence of different moisture forms**

*2.1.1. Super-Repellent CPSC in rain conditions*

To investigate how super-repellent coating behaves as a moisture barrier for bulk water, we performed rain simulation experiments **(Figure 3, Video S4, S5 and S6)**. The rain simulation was done by continuous dropping of the droplets with a pipette from 2 cm height and the frequency of ~3 drops per second on the horizontally placed CPSCs with carbon electrodes facing up (back side) at 30-40% initial relative humidity (RH). Simulated sunlight is irradiated on the glass side (front side) of the CPSCs by the mirror reflection, and the intensity of the incident light was attenuated to approximately 0.4 Sun. Besides the light irradiation, the mirror was also used to simultaneously observe the moment of degradation in CPSCs (color changes to yellow). The PV performance of CPSCs during the rain simulation experiment was continuously monitored. The photocurrent density was recorded with Zahner potentiostat at the voltage corresponding to the maximum power point ($V_{MPP}$) under approximately 0.4 Sun.

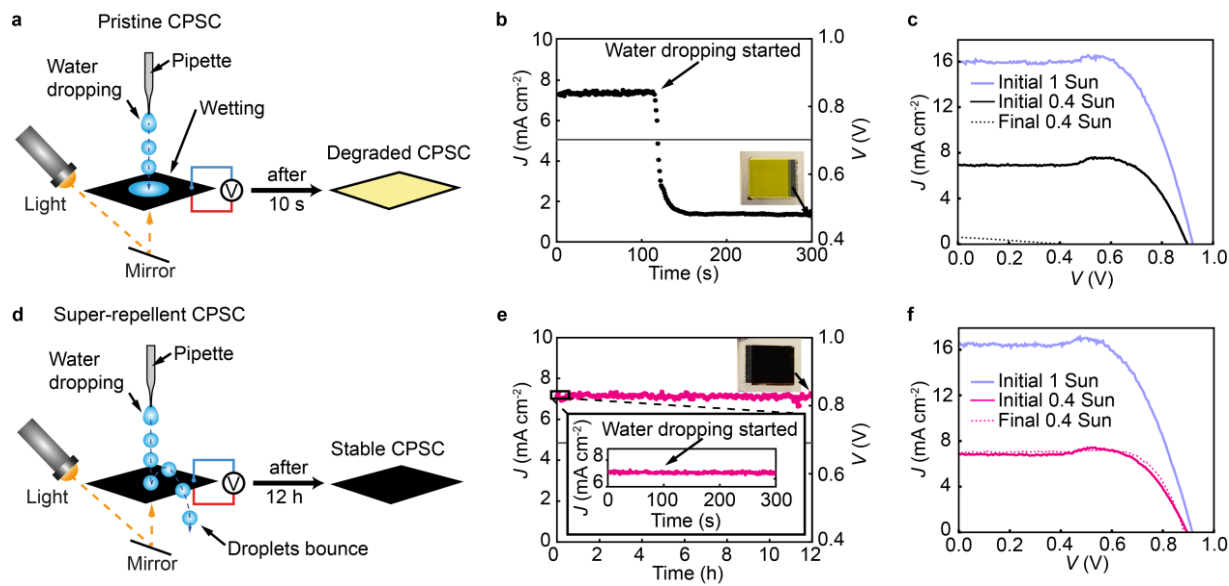

**Figure 3.** Rain simulation experiments. (a) Schematic illustration of the setup showing pipette for water dropping (rain simulation), light, and mirror for sunlight simulation. Due to strong wetting, the pristine CPSC degraded in 10 s after water dropping started. The P-V parameters were measured simultaneously. (b) Current density vs time for pristine CPSC under illumination (0.4 Sun) and bias 0.695 V ($V_{MPP}$). (c) J-V curves of the pristine CPSC at the initial 1 Sun and 0.4 Sun (attenuated due to the reflection through a mirror) and after the water dropping test under 0.4 Sun. (d) Schematic illustration of the setup for super-repellent CPSC showing droplet bouncing mechanism due to high repellency resulting in stable PCSC even after 12 h. (e) Current density vs time of super-repellent CPSC under illumination (0.4 Sun) and bias 0.67 V ($V_{MPP}$). (f) J-V curves of super-repellent CPSC with the same parameters from (c). The active area was 0.64 cm$^2$ (achieved by using a mask).

The current density of pristine CPSC sharply decreased in only 10 s after exposure to macroscopic water droplets (Figure 3a, b), which matched the rapid color change from black to yellow (Figure S5 and Video S4). This indicates quick decomposition of the perovskite absorber, resulting in cell damage and declined performance (Figure 3c and Table S4). In contrast, the current density of super-repellent CPSC stays remarkably stable at the maximum power point voltage ($V_{MPP}$) throughout 12 hours of exposure to macroscopic water droplets (Figure 3d, e) without degradation in cell appearance (Video S5) and PV performance (Figure 3f and Table S4). The superior performance of super-repellent CPSC is due to a low CAH and small friction forces that enable highly mobile droplets. The droplets thus easily bounce off the super-repellent CPSC (Video S6), reducing the contact time (between droplet and CPSC) and protecting the CPSC from water-induced damage. Thus, a thin layer of super-repellent coating serves as a remarkably stable shield under the impact of ~130 000 droplets and blocks the mass transfer of water.

*2.1.2. Super-Repellent CPSC in a humid environment*

To study the vapor-induced degradation in a humid environment, we performed the dark storage aging test at room temperature (RT) and relative humidity (RH) 65% and 85%. One set of 12 unencapsulated CPSCs (Batch I, 6 pristine CPSCs, and 6 super-repellent CPSCs), was placed in an automated environmental chamber (VCL 4006, Vötsch Industrietechnik, Germany) at the RT (23˚C) and the RH of 65 % and 85 % for specific durations (Figure 4). The carbon electrode was facing up to fully expose CPSCs to the humid environment during the aging test, and PV parameters were monitored throughout 1029 hours (Figure 4).

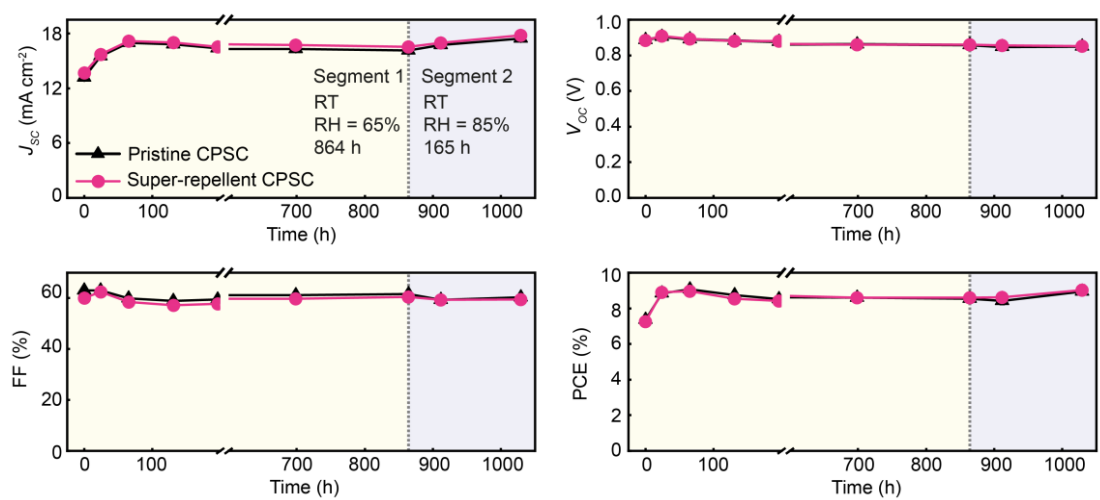

**Figure 4.** Dark storage test at room temperature (RT) and relative humidity (RH) 65% and 85%. Evolution of average (of forward and reverse scans) PV parameters of unencapsulated pristine and super-repellent CPSCs (6 cells in each group) under 1 Sun (aperture area of 0.64 cm$^2$). The aging test consists of two segments: segment 1 was conducted at RT and 65% RH for 864 hours, and segment 2 was conducted at RT, 85% RH for 165 hours. The distribution of PV parameters after each segment is reported in Figure S6.

The initial average PCE values for pristine and super-repellent CPSCs were almost identical (pristine CPSCs 7.4 ± 0.1 % and super-repellent CPSCs 7.3 ± 0.1 %) as well as the other PV parameters (Table S5, Supporting Information) confirming excellent integration of super-repellent coating. The overall behavior of these CPSCs was observed for ~ 43 days. In the first 65 h (of denoted segment 1, Figure 4), the average PCE significantly increased by 23% compared to the initial values (Figure 4 and Table S6). This was followed by a slight decrease in average PCE (during 864 hours) and then with its final increase (after 864 hours, end of segment 1) of 15.9 % ± 2.8 % (pristine CPSCs) and 18.5 % ± 1.7 % (super-repellent CPSCs) compared to initial PCE (Figure 4, and Table S6). The overall PCE improvement is mainly due to increased photocurrent (the open circuit voltage and fill factor remain stable).

Interestingly, with increasing RH to 85% in segment 2 (Figure 4), the photocurrent and PCE of both pristine and super-repellent CPSCs recovered (Figure 4). After 165 hours of segment 2, the average PCE increased (compared to the initial value) by 21.5 % ± 3.2 % for pristine CPSCs and 24.5 % ± 3.4 % for super-repellent CPSCs (Table S6). Surprisingly, the enhanced PCE at relatively high humidity (i.e., 65 % and 85 %) was maintained during 1029 h, not only for super-repellent CPSCs but also for unencapsulated pristine CPSCs. The enhancement in photocurrent and PCE for pristine CPSCs can be attributed to the interaction with humidity.[53–57] In CPSCs, the hydrophobic carbon electrode can block the large water droplets that might be produced under high-humidity conditions, and only allows the gaseous vapor to pass, enabling perovskite crystal growth without causing damage to its chemical structure.[53] This also further enhances the interface between perovskite and different layers of the printed stack.[53,57] Since the same was observed for liquid-repellent CPSCs (same evolution behavior, no statistically significant differences in the average change percentage of PV parameters, except for the Fill Factor (FF), Table S7), we can conclude that the eventual performance enhancement of super-repellent PCSCs is mainly due to the porous structure of the super-repellent coating (Figure 1f). Contrary to the bulk water (**Figure 5a**), where the water droplets are repelled (due to low CAH and friction forces, droplets bounce off the CPSCs), in a humid environment, it allows the water vapor to penetrate through the pores (Figure 5b). It thus absorbs a small vapor concentration (physisorption), resulting in perovskite crystal growth with preferential orientation and increasing the efficiency of CPSCs.

*2.1.3. Super-Repellent CPSC in supersaturating vapor*

To study the stability of super-repellent CPSCs in supersaturating vapor, we performed an aging test at elevated temperature on the same CPSCs from the previous aging test in the environmental chamber. The CPSCs were placed in the chamber with RH of 85 % and RT, and the temperature was increased from 23 °C to 40 °C (keeping constant RH). At the beginning of the experiment, we noticed water condensation on the carbon electrode due to reached supersaturation (supersaturated air with RH >100%). The supersaturation happened due to the temperature difference between the CPSCs and the environmental chamber.[58] When the air warms up to $T_{\text{air}} = 40°C$ (during the time $\Delta t_{\text{air}}$), there is a delay in warming up of CPSC ($\Delta t_{\text{CPSC}} > \Delta t_{\text{air}}$) due to the low heat transfer rate[59] of air and the larger heat capacity[59] of the CPSC (solid) compared to air (Figure 5f). Thus, the sample temperature falls behind the air ($\Delta T = T_{\text{air}} - T_{\text{CPSC}}$), and consequently, the humidity in the sample increases. This becomes more significant going from the sample surface to its depth (the core of the sample is colder than its surface during the warming-up time). Due to this, the RH at the CPSC reaches a value higher than 100%, and macroscopic condensation occurs, which is more severe inside the CPSC than on its surface. This results in significant liquid water condensation inside the CPSC, causing chemisorption and degradation of CPSC (color changes to yellow). When the sample finally equilibrates to the air temperature (40 °C) and humidity 85%, due to adsorption hysteresis (Figure 5f), the amount of condensed water inside the sample does not decrease to the same amount during physisorption (water absorption in the warm-up time).[60] This was confirmed with J-V measurements after 48h of exposure to 40 °C and RH of 85%. The pristine and super-repellent CPSCs show degradation in PCE of 61.7 % ± 10.0 % and 55.2 % ± 15.9 %, respectively, compared to the values before the test (Figure S7 and Table S8-10). It is worth to note that the super-repellent coating is stable at elevated temperatures, verifying by imaging with an atomic force microscope (AFM) (Figure S8). The AFM imaging was done by heating/cooling cycles (65°C/RT, 85°C/RT, and 100°C/RT), showing coating roughness unchanged through all cycles (Figure S8). Nevertheless, the coating does not affect the degradation rate of the super-repellent CPSCs under high-temperature conditions. This was demonstrated by a dark storage test at 65 °C in ambient for another set of 12 CPSCs (Batch I, 6 pristine CPSCs, and 6 super-repellent CPSCs) using a thermal chamber (Memmert, Germany). After 981 h of testing, the PCE of encapsulated pristine and super-repellent CPSCs reduced 35.3 % ± 2.2 % and 33.8 % ± 2.9 % of the initial values, respectively, with no statistically significant differences in the average change percentage between the two types of cells (Figure S9 and Table S11-S13).

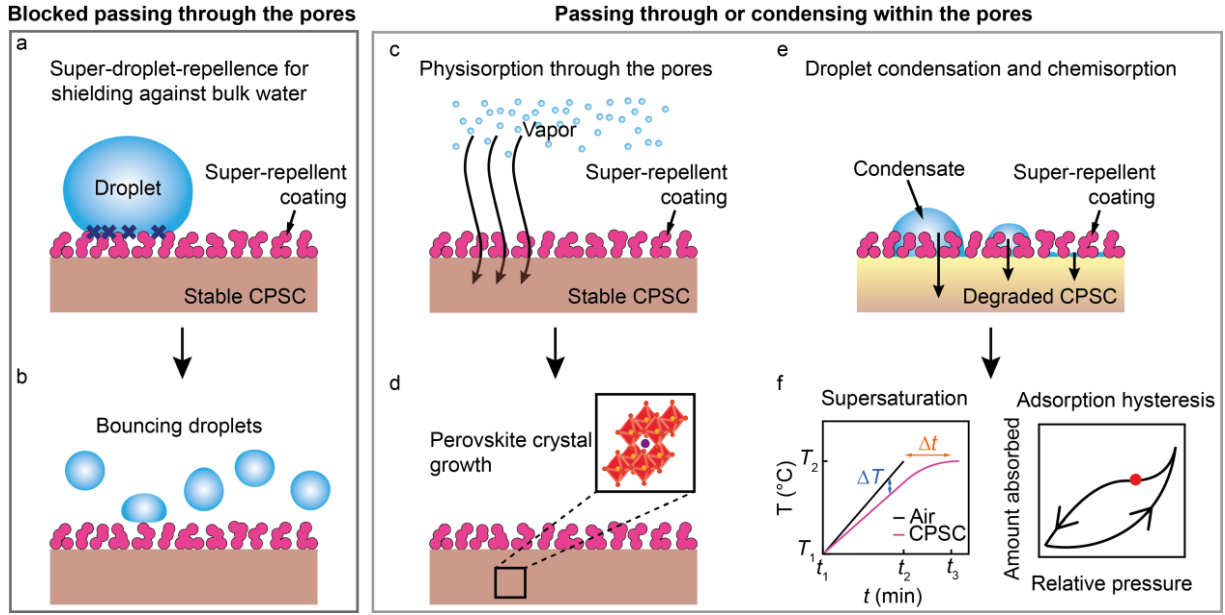

**Figure 5.** Repelling mechanisms for different moisture (water) forms. (a) Schematic illustration of shielding against bulk water showing blockage of mass transfer and resulting (b) droplet bouncing from the super-repellent CPSC. (c) Schematic illustration of vapor absorption through the pores of the coating resulting in (d) perovskite crystal growth. (e) Schematic illustration of condensation on super-repellent CPSC shows condensate growth within (or even below) the coating. (f) The graphs for supersaturation conditions and adsorption hysteresis characteristic of condensation from supersaturated vapor. The supersaturation graph shows the warm-up curve for air and CPSC indicated time delay $\Delta t$ of CPSC to reach the air temperature and temperature difference $\Delta T$ at which supersaturation occurs. After the CPSC equilibrate to air temperatures ($\Delta T = 0$) and RH 85%, due to adsorption hysteresis the amount of condensed water inside the sample does not decrease to the same amount during adsorbing water in the warm-up time (e.g., stays at the position indicated with the red dot).

To validate this further, we performed condensation tests (Batch II, 6 pristine, and 6 super-repellent CPSCs) in a homemade humidity chamber (Figure S10). After the supersaturation, the droplets condensed on the CPSCs (approximately after 30 min, Figure S10), and we continued condensation studies for 90 minutes more (2 hours total). There was a clear difference in the density of condensed droplets and their sizes between pristine and super-repellent CPSCs (Figure S10). The droplets formed on super-repellent CPSCs were smaller, less spread, and less dense compared to pristine CPSCs, which is attributed to the higher $\theta_{\text{avd}}$. As explained earlier, the coating could not prevent the CPSCs from degradation as expected for thin transparent nanoporous super-repellent coatings. Even in the case of repellent material, it is expected to fail in fogging experiments due to its porosity, as the tiny condensed droplets form within or below the coating. As droplets grow, they form sticky droplets that do not have enough energy to jump off the coating (Figure S10). These smaller condensates further serve as favorable points to accumulate water and behave like hydrophilic material, similar to pristine CPSCs (Figure 1b). Similar phenomena were observed on tilted surfaces of the pristine and super-repellent glass slides (Figure S11). While the average PCE of super-repellent CPSCs dropped to $0.1 \pm 0.1$ % (for pristine CPSCs to $0.4 \pm 0.5$ %) (Figure S12, Table S14), the super-repellent coating remained stable after the condensation test and elevated temperatures (was not damaged) as it was able to repel droplets (Video S7).

## 3. Conclusion

In summary, we focused on understanding the moisture-preventing mechanisms of thin nanostructures super-repellent coating from different moisture forms (falling water droplets vs water vapor vs condensed water droplets). To tackle this, we successfully integrated super-repellent coating (commercially available Glaco) into carbon-based printable perovskite solar cells (CPSCs) and tested its performance in conditions that resemble the outdoor environment: rain conditions, humid environment, and supersaturating vapor. The coating demonstrated remarkable success in repelling water droplets, effectively preventing simulated rain-induced damage and contributing to the longevity of the CPSCs. The coating allows for the physisorption of water vapor in humid environments, enhancing the performance of super-repellent CPSCs (similar to pristine CPSCs). In supersaturating vapor, the condensation occurred within and below the coating due to the temperature difference between the CPSC and air in the environment, followed by the chemisorption and degradation of CPSCs. Thus, rain simulation experiments and condensation tests are crucial aging tests for future hydrophobic/superhydrophobic coatings development as they provide a key understanding of the interaction mechanisms of different forms of moisture with coating. The findings are significant for the field as implementing hydrophobic and superhydrophobic (water-repellent) coatings as a moisture-preventing barrier is becoming an emerging research direction. We anticipate our findings inspire the development of repellent coatings that prevent the unwanted water condensation and chemisorption.

## 4. Experimental Section

*Materials*

Fluorine-doped tin oxide (FTO)-coated glass substrate (Product code: TCO22-7/LI), Ti-Nanoxide T165/SP, Zr-Nanoxide ZT/SP, and Elcocarb B/SP and perovskite precursor solution containing methylammonium lead iodide ($MAPbI_3$) and 5-ammonium valeric acid iodide (5-AVAI) was purchased from Solaronix, Switzerland. Titanium diisopropoxide bis (acetylacetonate, 75 % in isopropanol) was purchased from Sigma-Aldrich. The super-repellent coating is Glaco Mirror Coat Zero (SOFT99))..

*Device fabrication*

*Unencapsulated devices.* Fluorine-doped tin oxide (FTO)-coated glass substrate (10 cm x 10 cm, 7 $\Omega$ Sq$^{-1}$) was etched with an automated fiber laser to fabricate individual cell electrodes followed with ultrasonic cleaning in 1% Hellmanex aqueous solution, acetone, and isopropanol solvents (15 minutes each). A thin (30-50 nm) compact TiO$_2$ layer (c- TiO$_2$) was then deposited on an active area (through a glass mask) by aerosol spray pyrolysis at 450 °C using titanium diisopropoxide bis (acetylacetonate, 75 % in isopropanol) dissolved in absolute ethanol (1:80 v/v) as precursor and oxygen as carrier gas. After cooling to room temperature (RT), mesoporous TiO$_2$ (400-600 nm), the insulating mesoporous ZrO$_2$ (1-2 μm), and the conductive porous carbon (10-12 μm) layers were deposited via sequentially screen-printing Ti-Nanoxide T165/SP, Zr-Nanoxide ZT/SP, and Elcocarb B/SP pastes layer by layer. After each screen-printing step, the printed layers were dried at 150 °C for 5 minutes before sintering. Sintering temperatures for mp-TiO$_2$ and mp-ZrO$_2$ layers were 500 °C for 30 minutes, while the porous carbon layer was 400 °C for 30 minutes to form the mesoporous triple-layer-based scaffold. After final sintering, the printed layers were cooled to RT. The perovskite precursor solution (Solaronix) was prepared by mixing lead iodide (1.2 M), methylammonium iodide (1.2 M), 5-ammonium valeric acid iodide (5% mol) and dissolving in gamma-butyrolactone and ethanol mixture (85:15 v/v). The solution was stirred on a preheated (70 °C) hot plate for 30 minutes. The warm, clear yellow solution was filtered through a 0.2 μm PTFE filter and cooled to room temperature. The solution was then deposited using a programmable multi-channel pipetting robot (Solaronix) on an as-fabricated substrate masked with polyimide cut-out shapes (Impregnation Masks, Solaronix). The wet substrates were allowed to dwell for several minutes to let the liquid sip into the porous structure. The perovskite crystals in the porous electrode structure were achieved with final annealing in an oven at 55 °C for 90 mins. For encapsulation procedure, see Supporting Information.

The 50 PSCs were prepared as described above, of which 34 CPSCs are from Batch I, and 16 CPSCs are from Batch II.

*Super-repellent coating application*

The superhydrophobic coating was prepared using Glaco Mirror Coat Zero, a commercial product of a colloidal suspension of hydrophobic silica nanoparticles in isopropanol. To achieve an efficient super-repellent layer, the spin-coating of Glaco was done directly on the carbon electrode of CPSCs, masked with Kapton tape to avoid spreading the solution on the non-active area of device. 0.2 mL solution was deposited using a three-step sequential spin coating method: 2000 rpm (90 s), 4000 rpm (40 s), and 3000 rpm (60 s), followed by annealing on a preheated (70 °C) hotplate for 10 mins (one-layer coating). The procedure is repeated three times to ensure full coverage (three-layer coating).

*Device characterization*

*Photovoltaic measurements.* The current-voltage (J-V) curves of CPSCs were acquired using a Keithley 2401 Source Metter under simulated AM 1.5G sunlight at 1000 W/m$^2$ (1 Sun) irradiance generated by a Xenon-lamp-based solar simulator (Peccell Technologies, PEC-L01, Japan) with the intensity calibrated with a reference photovoltaic cell AK-300 (Konica Minolta, Japan). The scan range from -0.1 V to 1 V and scan rate 4.2 mVs$^{-1}$ were applied to measure the devices covered with masks with apertures of 0.14 cm$^2$ and 0.64 cm$^2$ to define active areas.

*Scanning electron microscopy (SEM).* The SEM imaging was done using field-emission scanning electron microscopy (FESEM, Zeiss ULTRA plus).

*XRD measurements.* The XRD measurements were carried out using the same procedures described in the previous report.[37] The XRD data was measured using a Rigaku Smartlab diffractometer with Cu anode and Ge (220) double bounce monochromator.

*UV/Vis spectroscopy.* The reflectance of the fabricated CPSCs was measured with an integrated sphere in a UV/Vis/NIR Spectrometer (Lambda 950, Perkin Elmer).

*Atomic Force Microscopy (AFM).* The surface morphology and stability of the samples under different temperatures were performed with a Bruker Multimode 8 Atomic Force Microscope (AFM) with a high-temperature stage for sample heating. The system was coupled with the Thermal Applications Controller (TAC) from Bruker to control the temperature of the heat stage and the probe. A pyramidal silicon probe with a cantilever spring constant of 0.3 N/m (HQ:CSC37/Al BS, MikroMasch) was used. The topography images were acquired with NanoScope software using ScanAsyst (Peak Force Tapping) imaging mode from Bruker. Images of 5 µm x 5 µm with 256 x 256 pixels were acquired at a scan rate of 1.0 Hz and a peak force frequency of 2 kHz. First, the measurements were performed at room temperature (RT), after which the stage and the probe were heated at 65 °C, 85 °C, and 100 °C. Before increasing the temperature, the system was set back to RT in each imaging cycle. For each temperature, a set of 5 images was acquired after at least 30 min for stabilization. Image processing and root mean square (RMS) roughness measurements were performed with Gwyddion software.

*Contact Angle Goniometry (CAG).* Contact angles were measured using a conventional optical tensiometer (Attension Theta, Biolin Scientific) following the established protocol [61]. Advancing contact angles were measured by placing a 0.22 µl droplet on the surface and increasing its volume to 20 µL, with 0.025 µL/s. Then, an additional 20 µL was added, and receding contact angles were measured by decreasing droplet volume with 0.025 µL/s. The data were analyzed using Young-Laplce fitting with OneAttension software, and reported advancing and receding contact angles were those when the droplet baseline increased and decreased, respectively.

*Oscillating Droplet Tribometery (ODT).* Friction forces for moving droplets on super-repellent CPSCs were measured using ODT [39,40,41] as follows. A 5 µL water-like ferrofluid droplet (0.2 vol.% nanoparticles) was placed on the CPSCs by pipette and brought into oscillations by moving the magnets in a sinusoidal fashion. The motion was captured with a Phantom V1610 high-speed camera at 1000 fps. The droplet position was tracked and fitted to the analytical solution for harmonic oscillations to obtain the friction force. For more details, see Supporting Information.

**Supporting Information**

Supporting Information is available from the Wiley Online Library or from the author.

**Acknowledgements**

This work was supported by the Jane and Aatos Erkko Foundation and Technology Industries of Finland Centennial Foundation for CAPRINT project funding (#2430354811), European Research Council grant ERC-2016-CoG (725513-SuperRepel) and Academy of Finland Center of Excellence Program (2022-2029) in Life-Inspired Hybrid Materials (LIBER) (#346109). G. S. L. acknowledges the support received from the Academy of Finland (#320090 and #317437). C. T. K. M. acknowledges Optoelectronics and Measurement Techniques (OPEM) Research Unit, University of Oulu, for permitting the use of the humidity chamber (Model VCL 4006, Vötsch Industrietechnik, Germany).

**Author's contributions**

C.M.T.K., J.H., R.H.A.R., S.G.H., and M.V. designed the experiments. C.M.T.K. conducted the characterization and aging tests (rain simulation, high humidity, and saturated vapor experiments), SEM measurements and analyzed the data. H.A.N. conducted UV/Vis spectroscopy, wetting characterization and data analysis. A.M.S. and G.S.L. carried out AFM measurements and data analysis. D.M. and S.N. fabricated all the devices. N.M. reviewed the first draft of the manuscript. S.G.H. and M.V. conceived the idea and performed initial experiments. C.M.T.K., J.H., and M.V. planned and wrote the manuscript. All authors contributed to data analysis and read and commented on the manuscript.

**Conflict of interest statement**

There are no conflicts to declare.

Super-repellent coating (advancing contact angle > 167˚ and contact angle hysteresis 7˚) is introduced for the first time to enhance the stability of carbon-based printable perovskite solar cells (CPSCs) under rain simulation. Super-repellent CPSC is prone to degradation from supersaturating vapor. Rain simulation and condensation tests for repellent and super-repellent CPSCs are proposed.

Cuc Thi Kim Mai*, Janne Halme*, Heikki A. Nurmi, Aldeliane M. da Silva, Gabriela S. Lorite, David Martineau, Stéphanie Narbey, Naeimeh Mozaffari, Robin H. A. Ras, Syed Ghufran Hashmi and, Maja Vuckovac*

ToC figure

Repelling mechanisms of super-repellent CPSCs for different moisture (water)

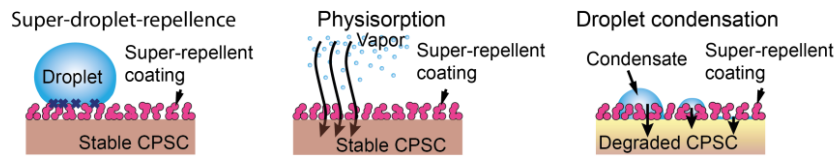